\documentclass[11pt,a4paper]{article}
\pdfoutput=1
\usepackage{jcappub}

\usepackage{amssymb,amsmath,amsthm}
\usepackage{braket}
\usepackage[T1]{fontenc}
\usepackage[latin1]{inputenc}
\usepackage[english]{babel}

\usepackage{float}
\usepackage{braket}
\usepackage{slashed}
\usepackage{commath}
\usepackage{diagbox}
\usepackage{color}
\usepackage{bm}
\usepackage{booktabs}
\usepackage{mathrsfs}
\usepackage{enumitem}
\usepackage{subfigure}

\renewcommand{\epsilon}{\varepsilon}

\newcommand{\p}{\partial}
\newcommand{\psibar}{\overline{\psi}}

\newcommand{\Ai}{\mathrm{Ai}}
\newcommand{\Bi}{\mathrm{Bi}}

\begin{document}
\title{Reheating in the kination epoch via multi-channel decay of gravitationally created massive scalars}

\author{Juho Lankinen,}
\author{Oskari Kerppo}
\author{and Iiro Vilja}

\affiliation{Turku Center for Quantum Physics, Department of Physics and Astronomy, University of Turku, Turku 20014, Finland}

\emailAdd{jumila@utu.fi}
\emailAdd{oeoker@utu.fi}
\emailAdd{vilja@utu.fi}

\abstract{
We provide a detailed study of reheating in the kination regime in a scenario where the particle content is produced by gravitational production of massive scalars decaying into massless scalars and fermions which eventually reheat the Universe. A detailed calculation is given by using Boltzmann equations and decay rates obtained using formalism of quantum field theory in curved spacetime. By numerical calculations the reheating temperature is found to be in the $10^{9}$-$10^{13}$ GeV regime. Moreover, the fermionic channel of decay is found to be the dominant channel of decay when the mass $m$ of the decaying particle is small raising the reheating temperature as opposed to a single scalar decay channel.
}

\maketitle

\section{Introduction} 
The inflationary scenario \citep{Guth:1981,Starobinsky:1980,Linde:1982,Sato:1980}, in which the Universe underwent a period of rapid expansion in the very early stages of its existence, is now a widely accepted paradigm in physics. The explosive nature of this expansion leaves behind a cold dark universe void of all matter, which must subsequently be created; the Universe must be reheated after inflation \citep{Bassett:2006,Allahverdi:2010}. Several methods for producing the necessary matter content has been investigated during the development of the theory; from the original collision of bubbles \citep{Guth:1981} to the now almost standard way of decay through oscillations of the inflaton \citep{Linde:1982}. Although these are perhaps the most widely used methods for reheating, ultimately one only needs a mechanism of creating radiation to eventually drive the expansion of the Universe.

A while ago it was realized that by using the result of gravitational particle creation, a feature of quantum field theory in curved spacetime, another mechanism for reheating the Universe might be realized \citep{Ford:1987,Spokoiny:1993}. As a consequence of this phenomenon, particles will be created by the very change of the metric itself as it changes from de Sitter metric into a postinflationary metric.  Some time ago, Ford considered a scenario where the Universe changes metric from de Sitter into either radiation- or matter-dominated era and found that gravitational particle creation is capable of reheating the Universe \citep{Ford:1987}. Spokoiny on the other hand considered an entirely different model altogether, one which is also capable of reheating the Universe. In Spokoiny's scenario the Universe ends up in a period he called deflation \citep{Spokoiny:1993}, nowadays  better known as kination \citep{Joyce:1997}, which can be realized when the kinetic energy of the inflaton dominates that of its potential energy. In this case the equation of state takes the form $\rho=p$ corresponding to a universe dominated by stiff matter.  While the idea of reheating by particle creation through a sudden or smooth change of the metric has received a lot of attention lately \citep{Akrami:2018,deHaro:2016a,Pallis:2006,Chun:2009,Salo:2017a,Hashiba:2019,Salo:2017b,Nakama:2019,Kunimitsu:2012,Herring:2020}, it is by no means the only mechanism capable of reheating: the very expansion itself creates particles making it possible to reheat the Universe using it \citep{Lankinen_Kerppo:2020}.

In our previous article \citep{Lankinen_Kerppo:2020} we investigated this type of reheating where  massive scalars gravitationally created in the stiff-matter era, decayed to massless scalars (radiation) which ultimately reheat the Universe. The standard model, or some other model, particles are eventually produced from these relativistic particles. It was found that this mechanism is capable of reheating the Universe to temperatures of about $10^4$-$10^{12}$ GeV. In the present article we will develop the model further and incorporate also a fermionic decay channel into the model. We therefore have two channels for the massive scalar to decay, fermionic and scalar, allowing us to explore a more general and realistic situation.

Two major ingredients of our study are the use of curved spacetime decay rates instead of the common Minkowskian decay rates and the use of the Boltzmann equations to describe the process. Describing particle decay in curved spacetime involves considerable conceptual and technical issues which make the generalization of normal in-out formalism from Minkowskian space non-applicable \citep{Audretsch:1986,Audretsch_Spangehl:1985,Lankinen_Malmi:2020,Buchbinder:1981}. We are therefore led to use the concept of added-up probability introduced in Ref. \citep{Audretsch_Spangehl:1985} for describing the decay in curved spacetime. This concept has recently been used in studying particle decay in spatially flat Robertson-Walker universes for a scalar and fermionic decay channels \citep{Lankinen_Vilja:2018b,Lankinen_Malmi:2020}. Although the Boltzmann equations are used in an integrated form, the model, together with the curved spacetime quantum field theoretic aspects, presents a novel approach to study reheating and also gives us a more precise picture of the whole process.

The paper is structured in the following way. In Sec. \ref{sec:2} we present our model and discuss about the assumptions made. The procedure for obtaining the reheating temperature is described in Sec. \ref{sec:3} and the results obtained in Sec. \ref{sec:4}. Finally, in Sec. \ref{sec:5} we discuss about the results.

\section{The Model}\label{sec:2}
In this section we will introduce all the necessary ingredients that make up our model and discuss the assumptions made regarding them.

\subsection{Background}
The Universe in the postinflationary scenario is described by a four-dimensional spatially flat Robertson-Walker metric
\begin{align}
ds^2=dt^2-a(t)^2d\mathbf{x}^2
\end{align}
given in standard coordinate time $t$ with a dimensionless scale factor $a(t)$. The scale factor is chosen as $a(t)=bt^n$ with $b$ a positive parameter controlling the expansion rate of the Universe and $n \in [0,1)$. The matter content of the Universe is assumed to be of the perfect fluid type characterized by a dimensionless parameter $\omega$ through $\rho=\omega p$, where $p$ is the pressure and $\rho$ the energy density of the fluid. The parameter $\omega$ is related to $n$ via the usual relation $n=2/(3\omega+3)$ \citep{Kolb_Turner}. The stiff matter is described by $\omega=1$, radiation by $\omega=1/3$ and ordinary matter approximated by $\omega\approx 0$ \citep{Kolb_Turner,Chavanis:2015}. Consequently, the Universe dominated by stiff matter is described by the parameter $n=1/3$, radiation-dominated Universe by $n=1/2$ and the matter-dominated Universe by $n=2/3$.  Since the scale factor is dimensionless, the parameter $b$ depends on $n$. This parameter appears in the Boltzmann equations only through the particle creation rate in the stiff-matter era so its units are fixed to be $\text{GeV}^{1/3}$.

We consider a massive scalar field $\phi$ with mass $m$ propagating in this spacetime which interacts with a massless scalar field $\chi$ and a massless Dirac spinor field $\psi$. The Lagrangian density is given by
\begin{align}
\mathcal{L}=\mathcal{L}_\phi+\mathcal{L}_\chi+\mathcal{L}_\psi,
\end{align}
where the massive scalar Lagrangian density is given by
\begin{align}
\mathcal{L}_\phi=\frac{\sqrt{-g}}{2}( \p_\mu \phi \p^\mu\phi-m^2\phi^2-\xi R\phi^2 ),
\end{align}
where $g$ is the determinant of the metric, $R$ the Ricci scalar and $\xi$ is the coupling to gravity. The value $\xi=1/6$ is known as conformal coupling in four dimensions and $\xi=0$ is the minimal coupling. The massless scalar is described by
\begin{align}
\mathcal{L}_\chi=\frac{\sqrt{-g}}{2}( \p_\mu \chi \p^\mu\chi-\frac{1}{6} R\chi^2)-\sqrt{-g} \lambda \phi \chi^2,
\end{align}
where $\lambda\neq 0$ is the scalar coupling constant. The spinors are incorporated into general relativity via the tetrad field $e_a{}^\mu$, where the latin indices refer to local inertial coordinates while greek indices refer to general coordinates. With this formalism, the fermion Lagrangian is given by
\begin{align}\label{eq:L_psi}
\mathcal{L}_\psi=\frac{i\sqrt{-g}}{2}( \psibar \gamma^\mu \nabla_\mu \psi-(\nabla_\mu \psibar)\gamma^\mu\psi)-\sqrt{-g} h \phi \psi\psibar,
\end{align}
where $h\neq 0$ is the fermionic coupling constant and $\psibar=\psi^\dagger\gamma^0$ the Dirac conjugate spinor. The curved space gamma matrices are defined via the tetrad as $\gamma^\mu=e_a{}^\mu \gamma^a$ satisfying the anticommutation relations $\{\gamma^\mu,\gamma^\nu \}=2g^{\mu\nu}$. The matrix $\gamma^a$ is the usual flat spacetime gamma matrix. The covariant derivative appearing in \eqref{eq:L_psi} is defined with the help of a spin connection $\Gamma_\mu$ as
\begin{align}
\nabla_\mu:=\partial_\mu+\Gamma_\mu,
\end{align}
where
\begin{align}
\Gamma_\mu=\frac{1}{8}[\gamma^a,\gamma^b]e_a{}^\nu \partial_\mu e_{b\nu}.
\end{align}

In the above, the massive $\phi$ particle is arbitrarily coupled to gravity while the massless fields are assumed to be conformally coupled. The reason for this lies in the added-up method which requires the decay products to be massless conformally coupled particles in order for the decay rate to make physical sense \citep{Audretsch_Spangehl:1985}. Moreover, the use of conformally coupled massless particles simplifies the model considerably because there is no gravitational creation of these particles by the expansion of spacetime \citep{Parker:1968,Parker:1969,Parker:1971} and therefore the only contribution to the relativistic energy density comes from the decaying particles.

\subsection{Decay in curved spacetime}
	Cosmological calculations most often use decay rates obtained in Minkowski spacetime \citep{Kolb_Turner,Liddle_Lyth}. The reason for this lies in the numerous conceptual issues and technical difficulties one encounters when treating particle interaction using the formalism of quantum field theory in curved spacetime \citep{Audretsch_Ruger_Spangehl:1987,Audretsch:1986,Audretsch_Spangehl:1986,Audretsch_Spangehl:1987}. Ultimately, however, the Minkowskian field theory is only an approximation and when curvature of spacetime cannot be neglected anymore the curved spacetime particle decay rates should be used. The problem of calculating decay rates in curved spacetime has attained a wealth of interest in recent times and curved space decay rates have been obtained for various processes \citep{Lankinen_Vilja:2017b,Lankinen_Vilja:2018a,Lankinen_Vilja:2018b,Lankinen_Malmi:2020,Boyanovsky,Boyanovsky:2020,Lotze:1989a,Lotze:1989b}. For our purposes, it is only necessary to recall the obtained differential decay rates for a massive scalar to decay into fermions or scalars in curved spacetime.

The differential decay rate for a massive scalar to decay into massless conformally coupled scalars was obtained in Ref. \citep{Lankinen_Vilja:2018b} for a general power-law expansion as
\begin{align}\label{eq:Diff_Chi}
\Gamma_\chi(t)=\frac{\lambda^2 t}{32}\big| H_{\alpha}^{(2)}(mt) \big|^2,
\end{align}
where $H_{\alpha}^{(2)}$ is the Hankel function of the second kind. For decay into massless fermions, it was found in Ref. \citep{Lankinen_Malmi:2020} that the differential decay rate is given by
	\begin{align}\label{eq:Diff_Psi}
	\Gamma_{\psi}(t)=\frac{h^2 t^n}{32}\bigg|\frac{d}{dt}\Big(t^{\frac{1-n}{2}}H^{(2)}_\alpha(mt)\Big)\bigg|^2.
	\end{align}
In both Eqs. \eqref{eq:Diff_Chi} and \eqref{eq:Diff_Psi} the index $\alpha$ of the Hankel function is given by
	\begin{align}\label{alpha}
	\alpha=\frac{\sqrt{(1-n)^2-4n(2n-1)(6\xi-1)}}{2},
	\end{align}	
where the index $n$ is the same $n$ as in the scale factor. 

\subsection{Boltzmann equations}
	With the necessary background reviewed, we move now to present and solve the integrated Boltzmann equations governing the evolution of the energy densities. We begin with the energy density $\rho_\phi$ of the massive $\phi$ particles for which the integrated Boltzmann equations read as
\begin{align}\label{eq:Boltzmann1}
\dot{\rho}_\phi(t)+3H(t)\rho_\phi(t)=-\Gamma_\chi(t)\rho_\phi(t)-\Gamma_\psi(t)\rho_\phi(t)+w_\phi(t),
\end{align} 
where $H=\dot{a}/a$ is the Hubble parameter. The second term on the left-hand side accounts for the dilution of the (nonrelativistic) particles, while the first and second terms on the right-hand side describe the decay into massless scalars and fermions, respectively. The last term $w_\phi$ is the contribution of the gravitationally created massive particles to the energy density. It is assumed that this creation occurs in the stiff-matter-dominated era only and after the transition it is zero. In our previous article \citep{Lankinen_Kerppo:2020} we derived this contribution using the results of particle creation in a stiff-matter-dominated Universe of Ref. \citep{Lankinen_Vilja:2017a} and found it to be
	\begin{align}\label{eq:Gamma_phi}
	w_\phi(t)=&\frac{3(mb)^{13/3}}{32b}t\big[\Ai(-(3mt/2)^{2/3})^2 +\Bi(-(3mt/2)^{2/3})^2],
	\end{align}	
where $\Ai$ and $\Bi$ refer to the Airy functions. We obtain a formal solution for the differential equation \eqref{eq:Boltzmann1} as
\begin{align}\label{eq:Rho_phi}
	\rho_\phi(t)=\frac{1}{a(t)^3}e^{-\int_{t_0}^t [\Gamma_\chi(t')+\Gamma_\psi(t')]dt'}\int_{t_0}^t a(t')^3w_\phi(t')e^{\int_{t_0}^{t'} [\Gamma_\chi(t'')+\Gamma_\psi(t'')]dt''}dt',
	\end{align}
where $t_0$ denotes the initial time taken to be the time when inflation ends. We have also assumed that the initial energy density $\rho_\phi(t_0)$ is zero because at the end of inflation the Universe is practically empty of matter. Even though there might be a contribution to the energy density from particles created by the change of the metric from de Sitter into stiff matter \citep{Hashiba:2019}, its contribution is negligible as long as the time when inflation ends is sufficiently far away from the spacetime singularity \citep{Lankinen_Kerppo:2020}.

The Boltzmann equations governing the evolution of the energy density for the relativistic particles $\chi$ and $\psi$, namely $\rho_{\mathrm{rel}}\equiv \rho_\psi+\rho_\chi$, reads as
\begin{align}\label{eq:Boltzmann2}
\dot{\rho}_{\mathrm{rel}}(t)+4H(t)\rho_{\mathrm{rel}}(t)=[\Gamma_\chi(t)+\Gamma_\psi(t)]\rho_\phi(t),
\end{align}
where the second term on the left side accounts for the dilution and the right-hand side expresses the creation of $\chi$ and $\psi$ particles as the massive scalar decays into these. The formal solution of \eqref{eq:Boltzmann2} is given as
\begin{align}\label{eq:Rho_rel}
	\rho_{\mathrm{rel}}(t)=\frac{1}{a(t)^4}\int_{t_0}^t [\Gamma_\chi(t')+\Gamma_\psi(t')]\rho_\phi(t')a(t')^4 dt'.
\end{align}
Here we have set the initial energy density $\rho_{\mathrm{rel}}(t_0)$ to zero reflecting the fact that no decay has occurred. Finally, the energy density of the background in the stiff-matter-dominated era is given by \citep{Chavanis:2015}
\begin{align}\label{eq:rhostiff}
	\rho_{\mathrm{stiff}}(t)=\frac{1}{24\pi G_N t^2},
\end{align}
where $G_N$ is Newton's gravitational constant.

\section{Reheating via gravitational particle production}\label{sec:3}
In this section we will outline the procedure for obtaining the reheating temperature. It should be noted that there exist two different type of scenarios which could occur under the model investigated. First, it may be that the energy density $\rho_\phi$ of the gravitationally created particles grows very slowly so that the energy density $\rho_{\mathrm{rel}}$ of the decay products dominates when equilibrium with the background energy density $\rho_{\mathrm{stiff}}$ is reached. In this case the Universe becomes radiation dominated at the transition. 

On the other hand, it may be that the energy density $\rho_{\mathrm{rel}}$ increases more slowly so that the energy density $\rho_\phi$ dominates at the equilibrium point. In this case the Universe ends up being temporarily matter-dominated and only later turns to radiation-dominated universe as the massive particles decay. We must consider these two possible scenarios separately.

\subsection{Matter-dominated era}
If the energy density of matter is equal to the energy density of the background, provided that the energy density of the massive particles is greater than that of the relativistic particles at that point, the Universe ends up as matter-dominated. In this case we start with the condition
	\begin{align}
	\rho_\phi=\rho_{\mathrm{stiff}}, \ \text{with the restriction} \ \rho_{\mathrm{rel}}<\rho_\phi.
	\end{align}
From this equality we obtain the time $t_{eq}$ at which the two energy densities are equal and we can calculate the energy density at this time. The evolution of the Universe is then described by ordinary Boltzmann equations in the matter-dominated era without any particle creation processes. The energy density after $t_{eq}$ is given as a solution of the ordinary Boltzmann equation,
	\begin{align}
	\rho_\phi^{mat}(t)=\Big(\frac{t_{eq}^{2/3}}{t^{2/3}}\Big)^3 e^{-f(t_{eq},t,2/3)}\times\rho_\phi(t_{eq}),
	\end{align}
where the superscript $mat$ indicates that this corresponds to the energy density in the matter-dominated era. The function $f$ is defined as
\begin{align}
f(t_0,t,n)=\int_{t_0}^t[\Gamma_\chi(t')+\Gamma_\psi(t')]dt',
\end{align}	 
where the index $n$ is implicit in the decay rates $\Gamma$ and labels the Universe matter content. In this case $n=2/3$ corresponds to matter-dominated Universe.
The massless particle energy density  $\rho_{\mathrm{rel}}^{mat}$ in the matter-dominated era is given by
	\begin{align}\label{eq:RhoChiMat}
	\rho_{\mathrm{rel}}^{mat}(t)=\Big(\frac{1}{t^{2/3}}\Big)^4 \int_{t_{eq}}^t[ \Gamma_\chi(t')+\Gamma_\psi(t')]\rho_\phi^{mat}(t')(t'^{2/3})^4dt'+\rho_{\mathrm{rel}}(t_{eq})\Big(\frac{t^{2/3}_{\rm{eq}}}{t^{2/3}} \Big)^4,
	\end{align}
where the last term describes the dilution of the initial energy density calculated at the equilibrium.	
The Universe continues to be matter-dominated until the energy densities of the massive and massless particles are equal. Once again the energy densities are then equated
	\begin{align}
	\rho_{\mathrm{rel}}^{mat}(t)=\rho_\phi^{mat}(t)
	\end{align}
to obtain time $\tau_{eq}$ when they are equal. The Universe transfers to radiation-dominated era and we can calculate the energy density at this time, which is then used as an initial condition. 
The energy densities in the radiation-dominated era are given by
	\begin{align}\label{eq:RhoChiRad}
	\rho_\phi^{rad}=\Big(\frac{\tau_{eq}^{1/2}}{t^{1/2}}\Big)^3 e^{-f(\tau_{eq},t,1/2)}\times\rho_\phi^{mat}(\tau_{eq})
	\end{align}
for the massive scalars and
	\begin{align}
	\rho_{\mathrm{rel}}^{rad}(t)=\Big(\frac{1}{t^{1/2}}\Big)^4 \int_{\tau_{eq}}^t [\Gamma_\chi(t')+\Gamma_\psi(t')]\rho_\phi^{rad}(t')(t'^{1/2})^4dt'+\rho_{\mathrm{rel}}^{mat}(\tau_{eq})\Big(\frac{\tau^{1/2}_{\rm{eq}}}{t^{1/2}} \Big)^4,
	\end{align}
for the massless scalars.	
	
The function $f$ now has $n=1/2$, corresponding to radiation-dominated era. The final step in the calculation scheme is the calculation of the reheating temperature. To obtain this, we maximize the function $\rho^{rad}_{\mathrm{rel}}$ with respect to the time $t$. This gives us the reheating time $t_{rh}$ and also the effective reheating temperature  $\frac{\pi^2}{30}g_*T_{rh}^4=\rho_{{\mathrm{rel}},max}^{rad}$. Because the numerical values of the constant term and degrees of freedom $g_*$ is of the order of unity when the fourth root is taken, we neglect these factors when obtaining the reheating temperature.

\subsection{Radiation-dominated era}
The other possible situation is when the energy density of the massless particles dominates the energy density of the massive particles when the equilibrium is reached with the background energy density $\rho_{\mathrm{stiff}}$. In this case
	\begin{align}
	\rho_{\mathrm{rel}}=\rho_{\mathrm{stiff}}, \ \text{with the restriction} \ \rho_{\mathrm{rel}}>\rho_\phi
	\end{align}
and the Universe ends up straight into radiation domination. The massive particle energy density is given as a solution of the Boltzmann equations
	\begin{align}
	\rho_\phi^{rad}(t)=\Big(\frac{t_{eq}^{1/2}}{t^{1/2}}\Big)^3 e^{-f(t_0,t_{eq},1/2)}\times\rho_\phi(t_{eq}),
	\end{align}
where the index $n=1/2$ in the function $f$ indicates that we are in the radiation-dominated Universe. The energy density for massless particles is given as
	\begin{align}
	\rho_{\mathrm{rel}}^{rad}(t)=\Big(\frac{1}{t^{1/2}}\Big)^4 \int_{t_{eq}}^t [\Gamma_\chi(t')+\Gamma_\psi(t')]\rho_\phi^{rad}(t')(t'^{1/2})^4dt'+\rho_{\mathrm{rel}}(t_{eq})\Big(\frac{t^{1/2}_{\rm{eq}}}{t^{1/2}} \Big)^4,
	\end{align}
where $t_{eq}$ is the time when the background stiff matter energy density $\rho_{\mathrm{stiff}}$ and the energy density of the massless particles $\rho_{\mathrm{rel}}$ is equal. The reheating time is obtained in the exactly same way by maximizing the radiation energy density.

\section{Numerical results}\label{sec:4}
	In the previous section we have established the procedure how to obtain the reheating temperature. The integrations must be performed numerically and for this we used PYTHON programming with Planck units where $G_N=1$. Using Planck units allows us to explore the widest possible range of parameters and after the calculations have been performed, the results can of course be expressed in terms of natural units straightforwardly

We are still required to fix a total of six parameters: the initial time $t_0$ marking the end of inflation, the mass $m$ of the decaying particle, the expansion parameter $b$ and the coupling constants $\lambda,\xi$ and $h$. We will first shortly discuss the values where these parameters were set and after that we take a look at the numerical results for the reheating temperature and discuss about the features the results presented.

\subsection{Fixing the parameters}
	The initial time $t_0$ was set to be the time when inflation ends. Although not exactly known, we fixed it to be $t_0=10^{11}$ corresponding to about $t_0\sim 10^{-32}$ sec, a value commonly found in the literature \citep{Kolb_Turner,Liddle_Lyth}. For the expansion parameter $b$ we ran the simulation with the range of $b\in [10^{-1},10^1]$. For the couplings constants $\lambda,\xi$ and $h$ the following values were given. The coupling $\lambda$, which was assumed to be small in order for the perturbative expansion in $\lambda/m$ to work, was fixed to run with the mass through the relation $\lambda=\gamma m$ with values $\gamma=10^{-1}, 10^{-2}, 10^{-3}$. Note that the ratio $\lambda/m$ is a dimensionless quantity. The gravitational coupling $\xi$ was chosen to be conformal because the gravitational particle creation rate in Ref. \citep{Lankinen_Vilja:2017a} was calculated for conformally coupled massive scalars.

	The parameter $h$, corresponding to the fermionic coupling, was fixed to be $h=10^{-14}$. This value allowed us to explore the widest mass range possible; for values smaller than $h=10^{-14}$ numerical errors were produced and for larger values the mass range was smaller. Moreover, if only one channel of fermionic decay is considered, it can be thought of as the singlet neutrino coupling. But since the coupling $h$ is dimensionless, we may consider multiple channels of decay and the parameter $h$ as an effective coupling. Hence, if we have multiple fermionic decay channels $\Gamma_{\psi_i}$, we may consider them as $\sum_i \Gamma_{\psi_i}=\sum_i g_i^2 h_i^2\tilde{\Gamma}_{\psi}$, where $g_i$ and $h_i$ denote the degrees of freedom of the particle and the coupling constant of the respective channels of decay. The quantity $\tilde{\Gamma}_{\psi}$ is defined as $\Gamma_\psi=h^2\tilde{\Gamma}_\psi$ with $\Gamma_\psi$ given by equation \eqref{eq:Diff_Psi}. We may then identify $h_{eff}^2=\sum_i g_i^2h_i^2$ and think of $h$ as an effective coupling. 

	Finally, the mass of the decaying particle was constrained to the widest range possible which did not produce errors in the numerical integrations. This range was found to be in the range of $10^{-14}$-$10^{-7}$ in Planck units which corresponds to about $10^5$-$10^{12}$ GeV providing a fairly large account of masses. The lower bound of the mass range seems to coincide with the value of the coupling $h$ in that for masses lower than the value of the coupling $h$, numerical errors were produced. The simulation was ran with the parameters fixed as described above and the plots were produced with a grid of $100\times 100$ points of the parameters $m$ and $b$. While discussing the results obtained, we will focus on contrasting the results with our previous work \citep{Lankinen_Kerppo:2020} where the same scenario was considered with a single scalaric decay channel.

\subsection{Reheating Temperature}

		\begin{figure}[t]
	\mbox{\subfigure[]{\includegraphics[scale=0.5]{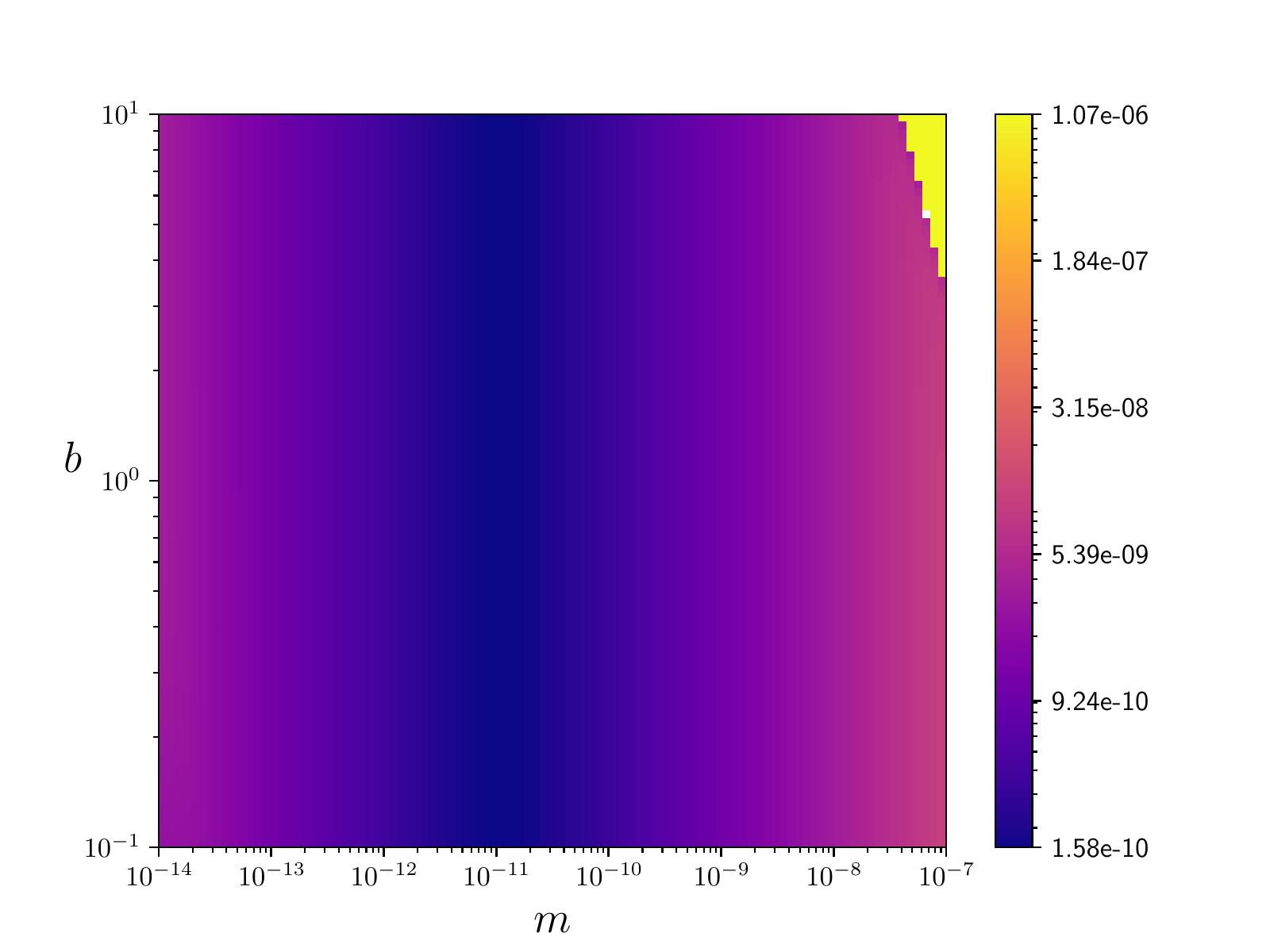}}\subfigure[]{\includegraphics[scale=0.5]{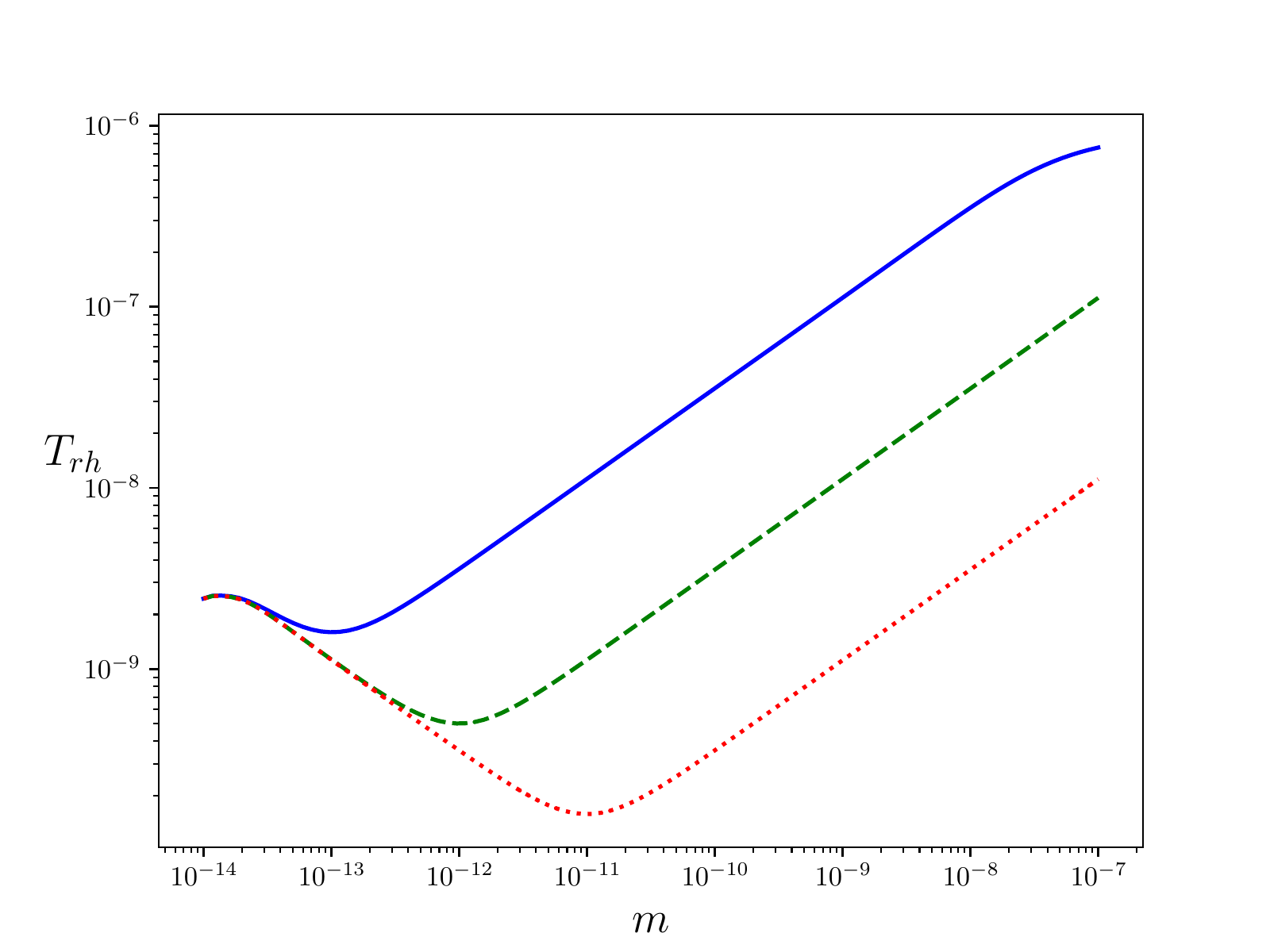}}}
	\caption{(a) Reheating temperature $T_{rh}$ as a function of mass $m$ and expansion parameter $b$ in Planck units with the values $\lambda/m=10^{-3}$, $h=10^{-14}$ and $\xi=1/6$. (b) Reheating temperature in Planck units as a function of mass $m$ with three values of the ratio $\lambda/m$; $10^{-1}$ (blue, solid), $10^{-2}$ (green, dashed) and $10^{-3}$ (red, dotted). The coupling is conformal, $h=10^{-14}$ and $b=0.1$.}\label{fig:Fig1} 
	\end{figure} 

	The reheating temperature $T_{rh}$ (figure \ref{fig:Fig1}a) was found to be practically independent of the parameter $b$ and to lie on the interval of about $10^{-10}$-$10^{-6}$ in Planck units depending on the parameters used. This corresponds to about $10^{9}$-$10^{13}$ GeV. Contrast to our previous work \citep{Lankinen_Kerppo:2020}, the lower bound of the reheating temperature is much higher. This is the result of the presence of a fermionic decay channel; the tendency for the scalar particle to decay into fermions is enhanced compared to the scalar channel \citep{Lankinen_Malmi:2020} which in turn raises the reheating temperature. The small area on the right upper corner in figure \ref{fig:Fig1}a correspond to the situation where the Universe ends up straight into the radiation-dominated era skipping the temporary matter dominated era.

	In figure \ref{fig:Fig1}b the reheating temperature is plotted with three values for the ratio $\lambda/m$ for $b=0.1$.
Comparing the three curves in figure \ref{fig:Fig1}b, we observe that the reheating temperature falls with the mass until at some point the fermionic channel of decay starts to dominate over the scalar channel raising the reheating temperature back up. If the decay channel is only to massless scalars, the reheating temperature would fall in a linear fashion as the mass decreases \citep{Lankinen_Kerppo:2020}. On the other hand the reheating temperature decreases in a linear fashion when the mass increases when only the fermionic channel is present, as was found in the simulation.  The value of the mass at which the fermionic channel starts to dominate depends largely on the values of both couplings $h$ and $\lambda$. For higher $\lambda$ it is seen from figure \ref{fig:Fig1}b that this value is smaller. The reason for this is that by increasing $\lambda$, decay into the scalar channel becomes more powerful thereby diminishing the effect of the fermionic channel because $\lambda$ is larger in comparison to $h$. We ran the simulation also for larger values of $h$. This had the effect of shifting the curves of figure \ref{fig:Fig1}b to the right while their shape remained the same; an increase of an order of magnitude in $h$ corresponds to a shift of about one order of magnitude in mass $m$ to the right.

	There is also a small bend on the left-hand side of figure \ref{fig:Fig1}b where the three curves meet. This is present only in the case where the expansion parameter $b$ is small, and for values larger than about $b=1$ the curves would jointly increase in a linearly fashion. This small bend can be attributed to the fact that the Universe ends up straight into radiation domination at these values. We will explain this transition into radiation domination in more detail in the next section, but here it sufficies to say that when $m$ and $b$ are small, there is less gravitational particle creation which decreases the energy densities and thereby the reheating temperature. At sufficiently low $b$ and $m$, the particle creation is so weak that the reheating temperature starts to decrease.

	Finally, we notice that the mass of the decaying particle is smaller than the reheating temperature i.e., $T_{rh}>m$ at least for the parameter range used. If thermal equilibrium is assumed, this would imply that the number density of $\phi$ quanta is higher than the number density of the relativistic particles in the thermal bath indicating that multiparticle processes may play a dominant role in order to decrease the number of particles. It is, however, known that the distribution of the $\phi$ particles or the decay products are not equilibrium distributions and the single-particle decay itself is not energy conserving \citep{Lankinen_Vilja:2017a,Lankinen_Vilja:2017b,Lankinen_Vilja:2018a,Lankinen_Vilja:2018b,Lankinen_Malmi:2020}, but the relativistic particles thermalize afterwards. Hence, when the system is out of equilibrium, the equilibrium description does not work and $T_{rh}>m$ would then not indicate the dominant role of multiparticle processes. For a more in depth look at this, we refer the reader to \citep{Lankinen_Kerppo:2020}.

\subsection{Transition times and different phases}	
		\begin{figure}[h]
	\centering
	\mbox{\subfigure[]{\includegraphics[scale=0.5]{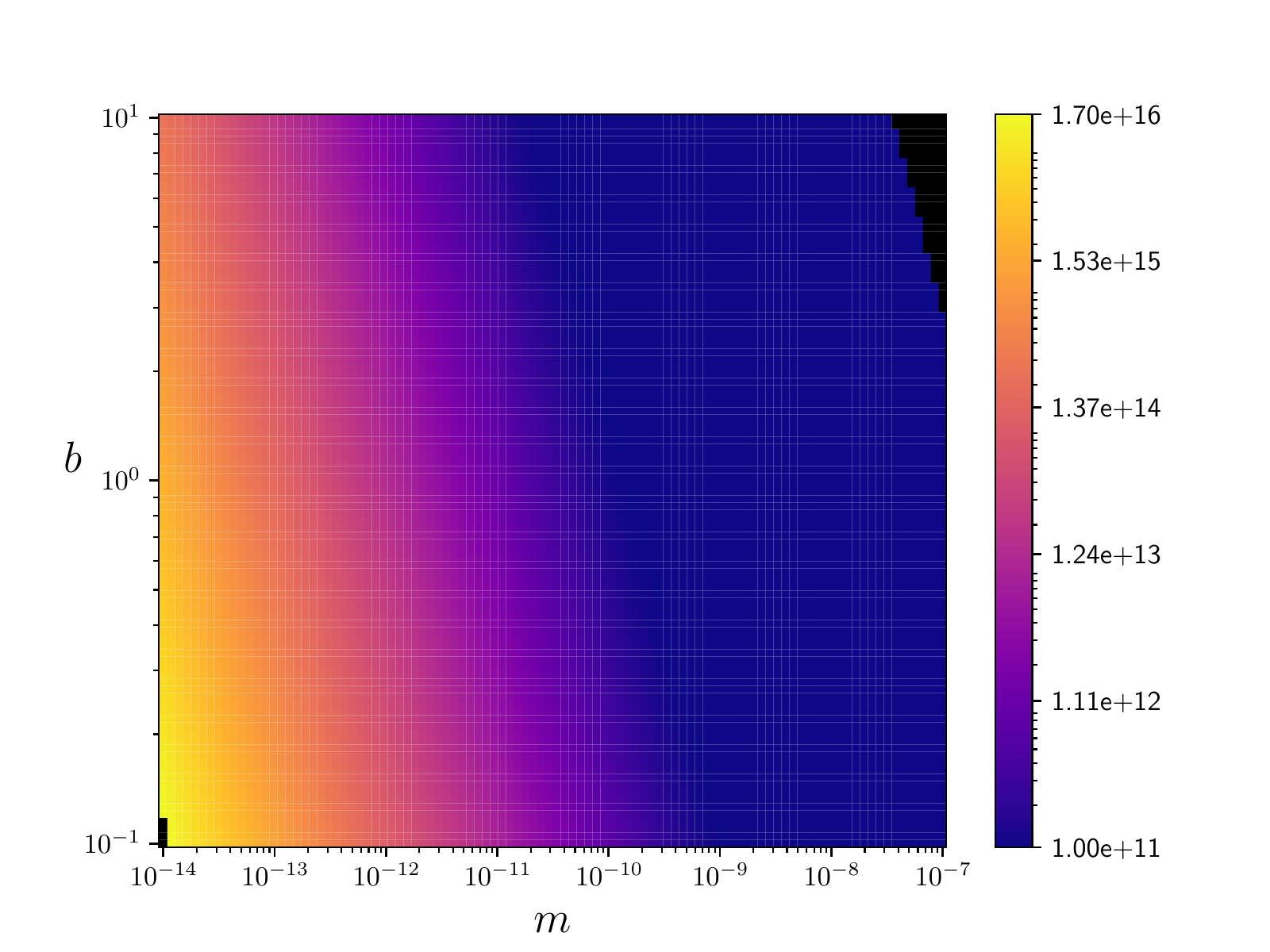}}\subfigure[]{\includegraphics[scale=0.5]{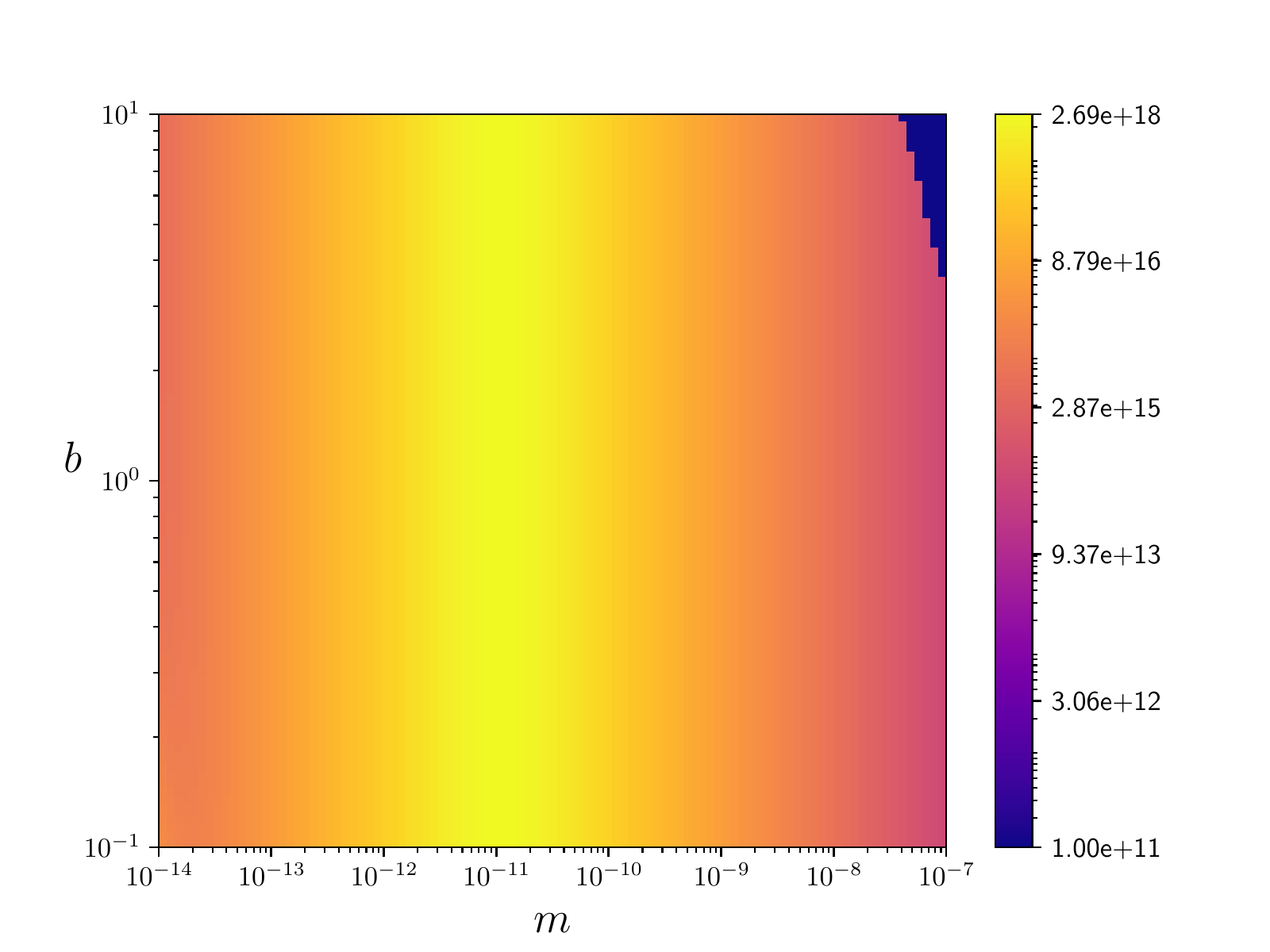}}}
	\caption{(a) Time of transition to a matter-dominated era as a function of mass $m$ and expansion parameter $b$ given in Planck units. The black areas in the upper right and lower left corners correspond to transition straight into radiation domination. (b) Reheating time as a function of mass $m$ and expansion parameter $b$ in Planck units. Both figures have the ratio $\lambda/m=10^{-3}$ and conformal coupling $\xi=1/6$.}\label{fig:Fig2} 
	\end{figure}

Taking a look at figure \ref{fig:Fig2}a, we consider first the transition into the temporary matter-dominated era. We see that the expansion parameter $b$ does have an effect on this time. In general, increasing the expansion parameter $b$ shortens the time it takes for the Universe to reach matter-dominated era as does increasing the mass $m$ of the decaying particle. These are reasonable to expect because the particle creation in a stiff-matter-dominated Universe is most effective when the mass $m$ and the parameter $b$ are large \citep{Lankinen_Vilja:2017a}. Hence, the particle creation is so explosive for these large values that the Universe reaches the temporary matter-dominated era almost instantly. Contrast to our previous work in Ref. \citep{Lankinen_Kerppo:2020} though, the time the Universe spends in this matter-dominated era is a few orders of magnitude shorter. This is  because in the present case there exist two different channels for the massive scalar to decay which decrease the energy density of these massive particles in this era more effectively. It seems reasonable to expect that the presence of more  channels of decay would shorten this time even further.

We would now like to discuss about the transition straight into the radiation-dominated era. The possibility of this scenario depends on the parameters $m$ and $b$ and can be attributed to particle creation in the stiff-matter era and particle decay in curved spacetime. The decay rate for the massive scalar to decay into massless scalars is the faster the more massive the particle is \citep{Lankinen_Vilja:2017b,Lankinen_Vilja:2018a,Lankinen_Vilja:2018b} while the situation is reversed for the fermionic channel for which the decay rate is faster the lighter the particle is \citep{Lankinen_Malmi:2020}. Hence, in the upper right corner of figure \ref{fig:Fig2}a, a large amount of particle creation takes place and since the decay into scalar channel is enhanced for large mass, the decay is so fast that radiation domination is almost immediately reached. As a consequence, the dominant particle species present in this case are bosons.

The radiation dominated portion in the lower left corner of figure \ref{fig:Fig2}a can also be explained by particle creation and decay rates. Since $m$ and $b$ are small, there are fewer particles created gravitationally and naturally it takes a longer time to reach equilibrium with the background. But as the decay into fermions is faster the smaller is the mass \citep{Lankinen_Malmi:2020}, they have enough time to decay and radiation domination is reached before equilibrium as $\rho_{\mathrm{rel}}$ increases. These notions are supported by figure \ref{fig:Fig2}b where it is seen that in this case the time it takes for the Universe to reheat is  quite long. If, on the other hand $b$ is large and $m$ is small, more particles are created and there simply is not enough time for these to decay until the equilibrium is reached. Hence the Universe ends up in matter-dominated period. Contrast to our previous study in \citep{Lankinen_Kerppo:2020}, this radiation-dominated phase does not exist for purely scalar decay channel and it occurs because the fermionic matter dominates the energy density.

\section{Discussion}\label{sec:5}
In this paper we have provided a novel approach to reheating in the kination epoch via the fermion and scalar decay channels using the Boltzmann equations. In this section we would like to discuss first about some of the assumptions made and then also provide more insight into the interesting features the two channel decay has provided.

With regarding to the assumptions, we have assumed that the Universe exhibits a sudden transition from the stiff-matter era either into radiation- or matter-dominated era. This is provided by the abrupt stop of particle creation after the change of phase has occured. It is possible that there exists some stiff matter after the transition, but as the background energy density scales as $\rho_{\mathrm{stiff}}\propto a^{-6}$, it is quickly diluted away and we do not therefore consider the particle creation from this residual stiff matter to be significant enough to affect our numerical results.
The second assumption concerns the added-up formalism used to derive the decay rates in curved spacetime. It is restrictive in the sense that the decay product particles are assumed to be massless and conformally coupled in order for the decay rate to make physical sense in curved spacetime \citep{Audretsch_Spangehl:1985}. Often one may consider this approximation of massless particles to be sensible when the decay product masses are much smaller than the decaying particle, i.e., $m_\psi,m_\chi \ll m_\phi,T$. This type of approximation must be taken with some caution though in curved spacetime because immediately as $m\neq 0$ the conformal invariance of the theory is broken and gravitational particle creation occurs. This in turn interferes severely with the mutual interaction of the process \citep{Audretsch_Spangehl:1985,Audretsch_Spangehl:1986,Audretsch_Spangehl:1987}. It is not clear if one may consider the decay products as having a negligible, but non-zero mass in this case. 



Finally, we ran the simulation also with a smaller $t_0$ to see how it affects the reheating temperature. From previous results it is known that the decay rate into fermions becomes more prominent as the spacetime singularity $t_0=0$ is approached \citep{Lankinen_Malmi:2020} and therefore we would expect to see the fermionic channel becoming more dominant as the time decreases. For the value of $t_0=10^{10}$, which was the smallest value that allowed a sensible range of parameters for calculation, we found no observable change in order of the reheating temperature or on the transition times although the numerical factors were changed. It can therefore be inferred, that the time $t_0$ should be very small, probably closer to the Planck time, in order for observable effects to take place.

In summary, we have in this article explored  reheating  in the kination epoch where the particle content is provided by gravitational particle creation during the expansion of the Universe. We used curved spacetime decay rates for a massive scalar to decay into radiation via scalar and fermionic channels. The reheating temperature was found to be in the $10^{9}$-$10^{13}$ GeV regime and to be independent of the expansion rate of the Universe. The results provide a deeper understanding into reheating occurring in the kination epoch but also increase our knowledge on the role that gravitational field has on the decay rates in curved spacetime.

\begin{acknowledgments}
J.L. would like to acknowledge the financial support from the University of Turku Graduate School (UTUGS). O.K. would like to acknowledge the financial support from the Turku University Foundation.
\end{acknowledgments}

\end{document}